\documentclass[10pt,final,doublecolumn]{IEEEtran}
\usepackage{subfigure}
\usepackage{amsmath}
\usepackage{latexsym}
\usepackage{graphicx}
\usepackage{bbding}
\usepackage{indentfirst}
\usepackage{setspace}
\usepackage{cite,balance}
\usepackage{float}
\usepackage{epstopdf}
\usepackage{amssymb}
\usepackage{amsfonts}
\usepackage{enumerate}
\usepackage{multicol}
\usepackage{color}
\usepackage{stfloats}
\usepackage{bm}
\usepackage{amssymb}
\usepackage{stfloats}
\usepackage{algorithm,algpseudocode,float}
\usepackage{lipsum}
\usepackage{makecell}
\usepackage{array}
\usepackage[table]{xcolor}
\usepackage{makecell}
\usepackage{pifont}
\IEEEoverridecommandlockouts
\allowdisplaybreaks[4]

\begin{document}
\title{Intelligent Reflecting Surface Aided Wireless Sensing: Applications  and Design Issues}
\author{{Xiaodan Shao,~\IEEEmembership{Member,~IEEE}, Changsheng You, \IEEEmembership{Member, IEEE}, and  Rui Zhang, \IEEEmembership{Fellow, IEEE}}
	 \thanks{X. Shao is with The Chinese University of Hong Kong, Shenzhen.}

\thanks{C. You is with the Southern University of Science and Technology.
}

\thanks{R. Zhang is with The Chinese University of Hong Kong, Shenzhen,  Shenzhen Research Institute of Big Data, and the National University of Singapore. }
}
\maketitle
\begin{abstract}
Intelligent reflecting surface (IRS) is an emerging technology that is able to significantly improve the performance of wireless communications, by smartly tuning signal reflections at a large number of passive reflecting elements. On the other hand, with ubiquitous wireless devices and ambient radio-frequency signals, wireless sensing has become a promising new application for the
next-generation/6G wireless networks. By synergizing low-cost IRS and fertile wireless sensing applications, this article proposes a new IRS-aided sensing paradigm for enhancing the performance of wireless sensing cost-effectively. First, we provide an overview of wireless sensing applications and the new opportunities of utilizing IRS for overcoming their performance limitations in practical scenarios. Next, we discuss IRS-aided sensing schemes based on three approaches, namely, passive sensing, semi-passive sensing, and active sensing. We compare their pros and cons in terms of performance, hardware cost and implementation complexity, and outline their main design issues including IRS deployment, channel acquisition and reflection design, as well as sensing algorithms. Finally, numerical results are presented to demonstrate the great potential of IRS for improving wireless sensing accuracy and the superior performance of IRS active sensing compared to other schemes.
\end{abstract}

\IEEEpeerreviewmaketitle

\section{Introduction}
\begin{figure*}[t]
\setlength{\abovecaptionskip}{-0.cm}
\setlength{\belowcaptionskip}{0.cm}
  \centering
\fbox{\includegraphics [width=0.90\textwidth] {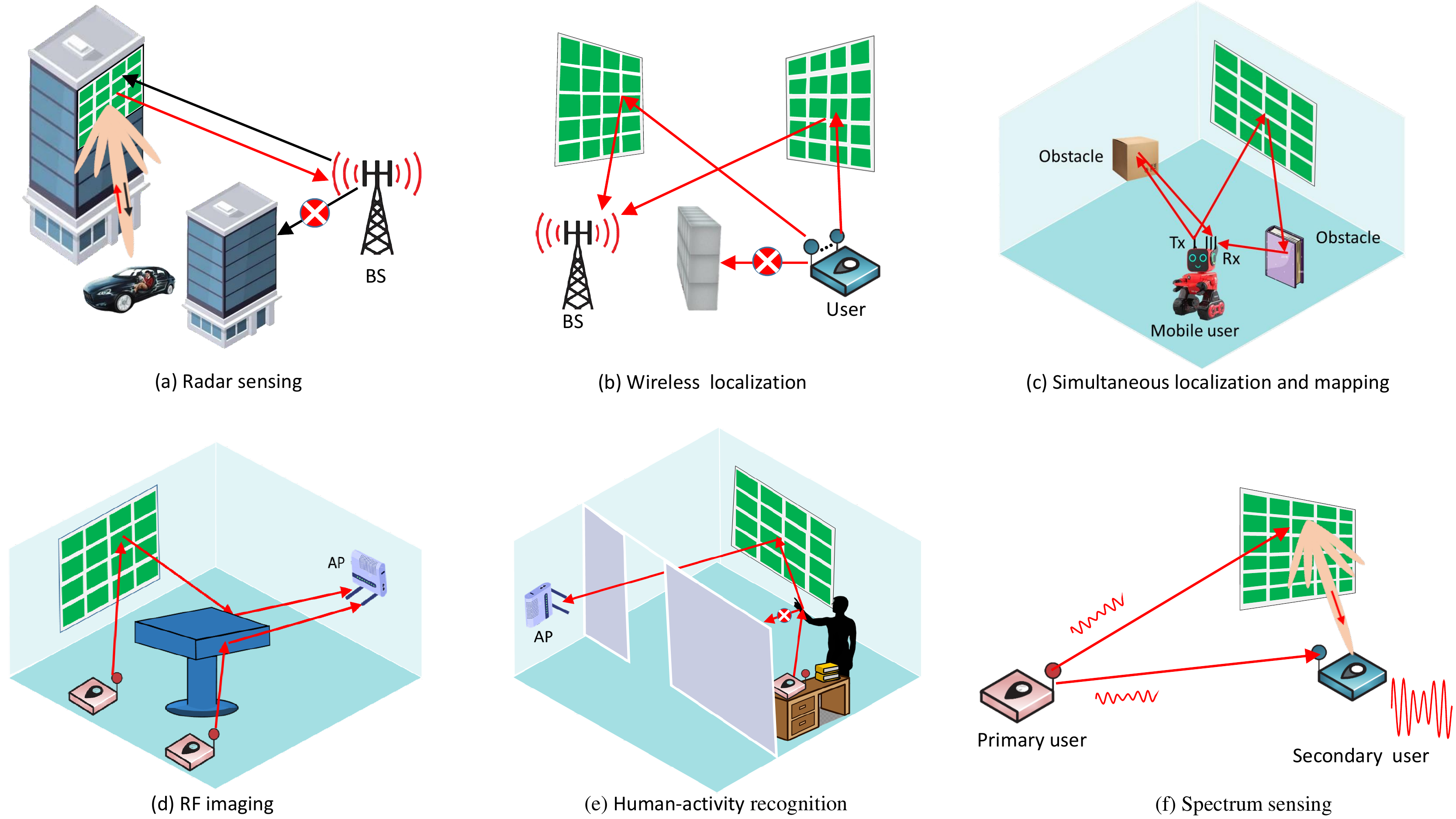}}
\caption{Typical wireless sensing applications and the use of IRS for improving their performance.}
\label{application}
\end{figure*}
\noindent The next-generation or 6G wireless networks will embrace a horizon of new applications such as tactile internet, autonomous transportation, ubiquitous sensing, extended reality and so on, which demand not only higher-quality wireless connectivity but also unprecedentedly higher-accuracy sensing than today's wireless systems \cite{6gisac}. Meanwhile, the widely deployed massive multiple-input multiple-output (MIMO) and wide-band
millimeter wave (mmWave) communication technologies have enabled high-resolution wireless signals in both space and time, which can be utilized to achieve high-performance sensing. Motivated by the above, 6G wireless networks are envisioned to share the communication platforms and resources such as base stations (BSs)/access points (APs), antennas, spectrum, waveforms, etc. for achieving the dual functions of communication and sensing cost-efficiently, leading to a new paradigm called \emph{integrated sensing and communication} (ISAC) \cite{ISAC1}.
Furthermore, wireless sensing can provide useful environment and device location/channel information, which can also be utilized efficiently for improving the communication performance \cite{liuan}.

Generally speaking, wireless/radio-frequency (RF) sensing refers to
detecting the presence or estimating various physical characteristics (e.g., position, orientation, speed, material) of objects in the environment
by exploiting radio wave transmission, reflection, diffraction, and/or scattering. Some typical applications for wireless sensing are given as follows: 1) {\emph{radar sensing}}, where the radar radiates electromagnetic waves to hit the target and the reflected echo signals are received to estimate the target's physical information such as distance, velocity, azimuth, height, etc.; 2) {\emph{wireless localization}}, where wireless device sends signal to one or more APs/BSs to estimate its location; 3) {\emph{simultaneous localization and mapping (SLAM)}}, which extends wireless localization by simultaneously estimating the location of the wireless device as well as scatterers/obstacles in its surrounding environment for constructing a map of
the unknown environment; 4) {\emph{RF imaging}}, which generates/reconstructs high-resolution images of environment objects by exploiting RF signals that bounce off the objects and are resolved in different dimensions such as time, space, and frequency \cite{imag}; 5) {\emph{human-activity recognition}}, which detects or recognizes human activities (e.g., gesture, sleep, and breathing) by leveraging the amplitude/phase variations in wireless signals; and 6) {\emph{spectrum sensing}}, which aims to obtain awareness about the spectrum usage and detect the existence of primary (high-priority) users to enable the communications of secondary (low-priority) users.

However, existing wireless sensing systems face critical challenges in practice, which fundamentally limit their performance in future wireless networks, elaborated as follows.
\begin{itemize}
\item[(1)] {\bf {Non-availability of line-of-sight (LoS) link:}}
    Wireless sensing performance can be significantly deteriorated in unfavorable propagation environment with severe signal blockages, such as radar sensing and human-activity recognition, as shown in Figs. \ref{application}(a) and 1(e). In these applications, only the signal that propagates over the LoS links is useful for sensing, while the signals from other non-LoS (NLoS) links cause interferences that can degrade the sensing accuracy.

    \item[(2)] {\bf{Lack of multipath propagation:}}
    In contrast, some sensing applications rely on the multipath propagation in the environment, such as wireless localization and SLAM, as shown in Figs. \ref{application}(b) and 1(c). Although more BSs/APs/devices can be deployed in the environment to create additional signal paths between the sensing transmitter and receiver, this entails higher cost and may not be practically feasible.

    \item[(3)]{\bf {Limited sensing range due to high path loss:}}
    Due to the increased path loss of wireless signals over distance, all sensing applications generally have limited operation range and/or accuracy. For example, in the case of spectrum sensing as shown in Fig. \ref{application}(f), the signal from the primary user may be severely attenuated when arriving at the secondary user, which makes its detection of the primary-user signal against the background noise practically difficult. Although the path loss can be compensated by employing massive MIMO receiver and/or increasing transmit power, these methods incur higher cost and more power consumption, thus may be infeasible when there are limited budgets on hardware and the power supply.
\end{itemize}

To address the above issues, \emph{intelligent reflecting surface} (IRS) or its various equivalents such as reconfigurable intelligent surface (RIS), which has been recently proposed to reconfigure the radio propagation environment for boosting the wireless communication capacity \cite{qingqing,overview}, can also be exploited to enhance the performance of existing wireless sensing systems. Specifically, IRS is a digitally-controlled metasurface consisting of a large number of passive reflecting elements, each being capable of inducing a controllable amplitude and/or phase change to the incident signal independently \cite{qingqing}. IRS possesses several appealing advantages for practical use. For example, it entails much lower hardware and energy cost as compared to the conventional active relay, since it is free of costly RF chains and amplifiers. Besides, IRS operates in a full-duplex mode without incurring processing delay and self-interference. Thus, integrating IRS into wireless sensing systems provides new opportunities to overcome their aforementioned performance limitations in a cost-effective way, which are further explained as follows.
\begin{itemize}
\item[(1)] {\bf {Create virtual LoS links:}} For example, in radar sensing (Fig. \ref{application}(a)), when environmental obstacle blocks the direct link between the sensing BS and target, IRS can be installed (e.g., on the facade of a high-rise building) to help establish a virtual LoS link from the BS to the IRS and then to the target for enabling the detection of the echo signal that is reflected back via the same virtual LoS link in the reverse direction \cite{RSPjoint, radarin}. Besides, for human-activity recognition (Fig. \ref{application}(e)), IRS can be used as a controllable reflector to create LoS path between the person's hand and a far-apart receiver to improve the recognition accuracy.

\item[(2)] {\bf {Establish controllable multipath propagation:}} For instance, in wireless localization (Fig. \ref{application}(b)), two (or more) IRSs can be employed as reflectors to create multipath propagation for enabling localization under blockage. For SLAM (Fig. \ref{application}(c)), IRS can create a multipath-richer propagation environment to more accurately estimate the user's location as well as the local map \cite{song}. Moreover, for RF imaging (Fig. \ref{application}(d)), to successfully image the backside of the object, IRS can be properly deployed and smartly controlled for its reflected beam to illuminate the object's blind spot.

\item[(3)]{\bf {Enable flexible passive beam scanning to compensate path loss:}} For example, in spectrum sensing (Fig. \ref{application}(f)), to increase the detection accuracy of the spectrum usage for the primary user, IRS beam scanning can be applied by dynamically changing reflected wave's focusing direction  for increasing its peak power towards the secondary user, hence boosting the primary-user signal strength received at the secondary user for more accurate detection \cite{specBei}.
\end{itemize}

The above new applications of IRS for different wireless sensing systems have a great potential to fundamentally improve their performance in practice. However, compared to IRS-aided communication,
IRS-aided sensing is still in its infancy and calls for great research efforts to resolve its unique challenges that differ significantly from their communication counterparts. Specifically, unlike
IRS-aided communication which targets for improving data transmission performance, IRS-aided sensing aims to accurately and efficiently detect, estimate, and extract useful information/features of targets subjected to environment noise and interference. Thus, new design issues pertinent to IRS-aided sensing arise, such as system architecture, IRS deployment and reflection design, sensing algorithm design, etc. This article is thus aimed to provide an overview of these important issues as well as propose promising approaches to tackle them in practical wireless sensing systems. For ease of exposition, we mainly focus on IRS-aided radar sensing in the sequel of this article, while similar discussions can be inferred for other wireless sensing applications as they usually face similar design issues and practical performance limitations, as illustrated in the above. The rest of this article is organized as follows. Section II
introduces various system architectures and discusses the main design issues for IRS-aided radar sensing. Section III presents numerical results to evaluate the performance of different IRS-aided radar sensing architectures and algorithms. In Section IV, other related research directions are discussed, followed by the conclusions made in Section V.

\section{IRS-Aided Radar Sensing: System Architectures and Design Issues}
In this section, we present the typical architectures and main design issues for IRS-aided radar sensing, including IRS deployment, channel acquisition, reflection design, and sensing algorithms. Promising approaches to resolve these issues as well as open problems worthy of further investigation are also highlighted.
\subsection{System Architecture}
\begin{figure*}[t]
\setlength{\abovecaptionskip}{-0.cm}
\setlength{\belowcaptionskip}{0.cm}
  \centering
\fbox{\includegraphics [width=0.85\textwidth] {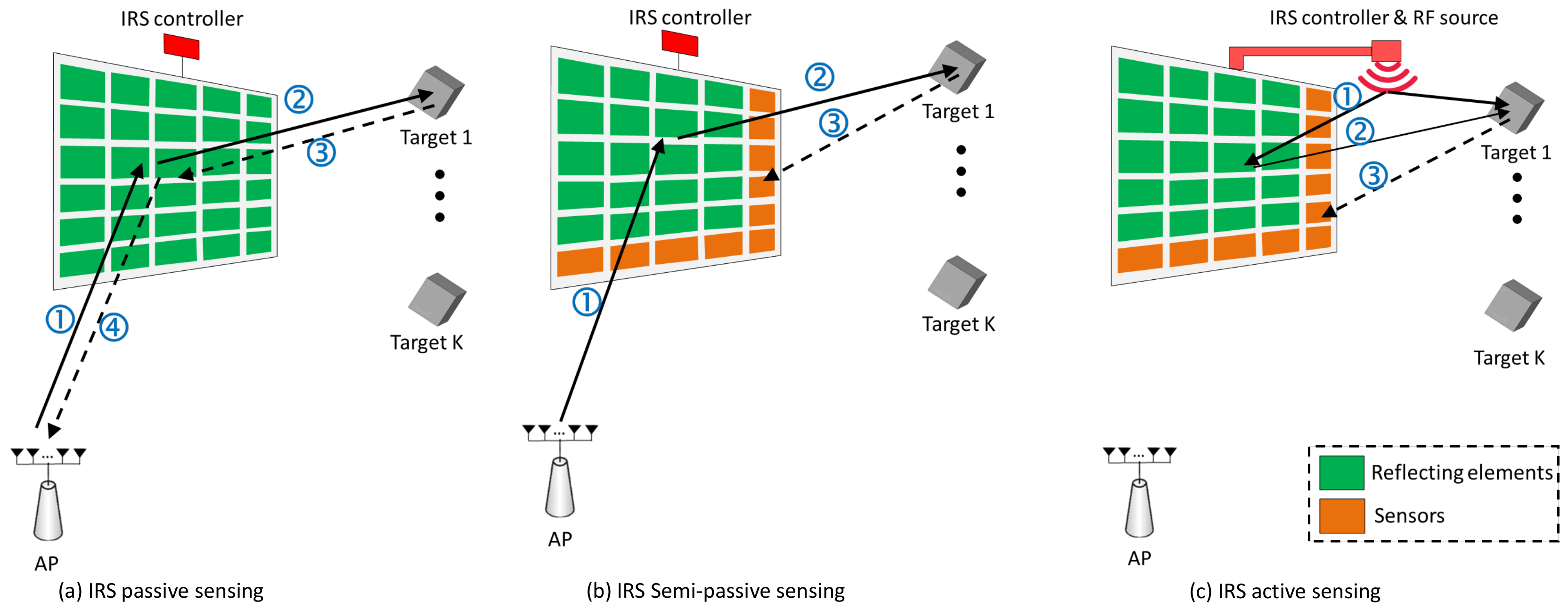}}
\caption{Schematics for three IRS-aided radar sensing architectures.}
\label{sensing}
\end{figure*}

Depending on whether receive/transmit RF chains are installed at the IRS or not, IRS-aided radar sensing systems can be generally classified into the following three categories: IRS passive sensing, IRS semi-passive sensing, and IRS active sensing. \emph{IRS passive sensing} is shown in Fig. \ref{sensing}(a), where the IRS consists of a number of passive reflecting elements and a smart controller only, and the controller is responsible for adjusting IRS reflections as well as exchanging low-rate information with the BS/AP. As such, passive IRS can only reflect signals from the BS/AP to illuminate the targets for detection. Although IRS passive sensing enjoys low hardware and energy cost, its sensing performance (e.g., accuracy and range) is practically limited. This is because the signal in IRS passive sensing needs to travel over a triple-reflection link from the BS's transmitter to its receiver, i.e., BS$\stackrel{\text{\ding{172}}}{\longrightarrow}$IRS elements$\stackrel{\text{\ding{173}}}{\longrightarrow}$target$\stackrel{\text{\ding{174}}}{\longrightarrow}$IRS elements$\stackrel{\text{\ding{175}}}{\longrightarrow}$BS. As a result, IRS passive sensing suffers severe product-distance path loss and hence low received signal-to-noise ratio (SNR).

The second category is \emph{IRS semi-passive sensing}, where additional low-cost sensors are installed on the IRS, as shown in Fig. \ref{sensing}(b). Thereby, the IRS can not only reflect signals from the BS, but also receive echo signals from the target directly for radar sensing. Compared with IRS passive sensing, IRS semi-passive sensing experiences less path loss, since its signal goes through a double-reflection link only, i.e., BS$\stackrel{\text{\ding{172}}}{\longrightarrow}$IRS elements$\stackrel{\text{\ding{173}}}{\longrightarrow}$target$\stackrel{\text{\ding{174}}}{\longrightarrow}$IRS sensors. However, if the sensing BS is far from the IRS, the sensing accuracy of IRS semi-passive sensing is still limited by the severe path loss in the BS-IRS link.

To address the above issues, a new type of IRS-aided radar sensing architecture, called \emph{IRS active sensing}, has been recently proposed \cite{shao_target} (see Fig. \ref{sensing}(c)). Specifically, in IRS active sensing, the IRS controller, conventionally responsible for IRS reflection control and information exchange only, also acts as a transmitter (equipped with an RF source) to send probing signals for sensing. In this case, the IRS controller placement is no longer flexible as in IRS passive/semi-passive sensing cases. Instead, it should be placed in front of IRS elements in close proximity, e.g., hanging out perpendicularly from the IRS front plane for the convenience of transmitting signals towards IRS reflecting elements, as illustrated in Fig. \ref{sensing}(c). Moreover, dedicated sensors are also installed on the IRS to receive echo signals from targets. As such, the IRS can actively send and receive radar signals for target localization by itself, without the need of any BSs/APs for signal transmission/reception as in conventional radar sensing systems. Although IRS active sensing generally incurs higher hardware and energy cost than IRS passive/semi-passive sensing, it is expected to achieve
superior sensing performance, since the IRS controller$\rightarrow$IRS elements link is much shorter than the BS$\rightarrow$IRS elements link in IRS passive/semi-passive sensing. Moreover, IRS active sensing can also leverage a direct echo link, i.e., IRS controller$\to$target$\to$IRS sensors, which can be combined constructively at the sensors with the IRS reflected echo link (i.e., IRS controller$\to$IRS elements$\to$target$\to$IRS sensors) by properly tuning the IRS reflection, thus further enhancing the sensing performance. In Table I, we summarize the comparison of the three IRS-aided radar sensing system architectures in different aspects.
\newcommand{\tabincell}[2]{\begin{tabular}
{@{}#1@{}}#2\end{tabular}}
\begin{table*}
\caption{Comparison of IRS-aided radar sensing architectures}
\centering
\begin{tabular}{>{\columncolor{blue!15}}c >{\columncolor{black! 10}}c >{\columncolor{blue!15}}c >{\columncolor{black! 10}}c >{\columncolor{blue!15}}c >{\columncolor{black! 10}}c >{\columncolor{blue!15}}c}
\bfseries \tabincell{c}{Architecture} &\bfseries Range & \bfseries {\tabincell{c}{Cost}}
 & \bfseries \tabincell{c}{Accuracy} &\bfseries \tabincell{c}{BS/AP needed?}
&\bfseries Path Loss
\\
\Xhline{1pt}
Active sensing &Long&High  &High
 &No&\tabincell{c}{Low }\\
\hline
\tabincell{c}{Semi-passive sensing} &Medium &Medium &Medium
&Yes
&\tabincell{c}{High }\\
\hline
Passive sensing &Short
 &Low
 &Low
 &Yes
 &\tabincell{c}{Very high}
\end{tabular}
\label{table}
\end{table*}

\subsection{IRS Deployment}
\begin{figure*}[t]
\setlength{\abovecaptionskip}{-0.cm}
\setlength{\belowcaptionskip}{0.cm}
  \centering
\fbox{\includegraphics [width=0.99\textwidth] {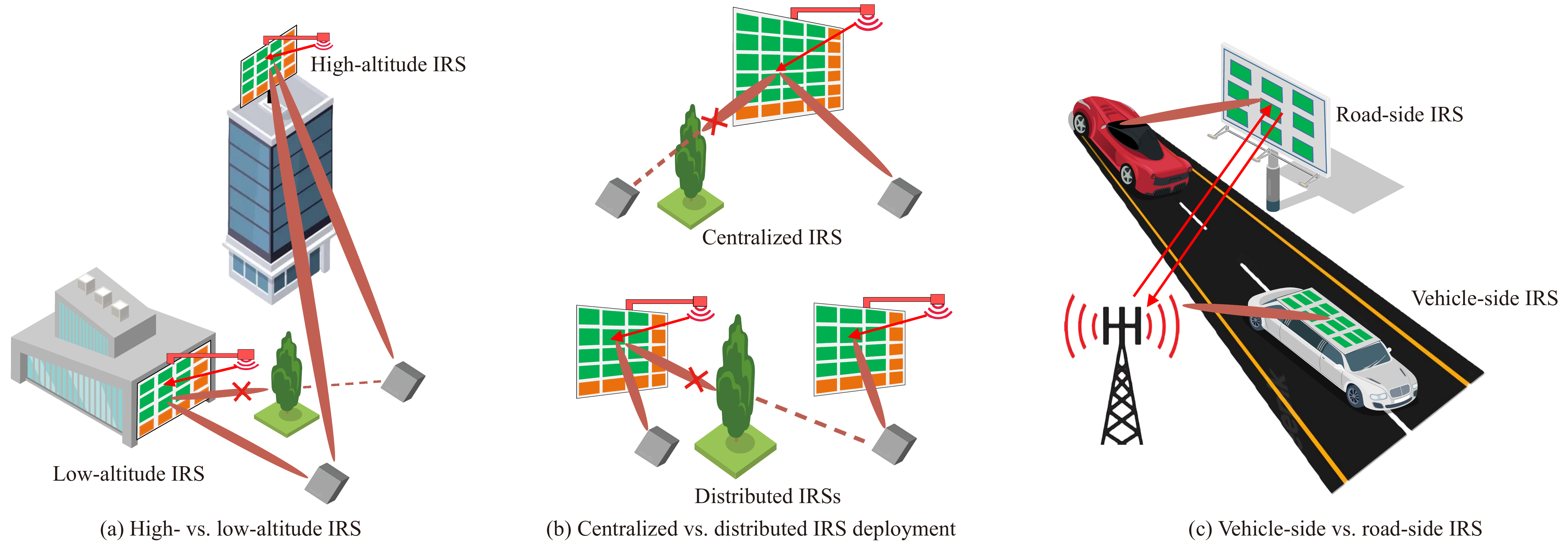}}
\caption{Different IRS deployment strategies in radar sensing system.}
\label{deployment}
\end{figure*}

How to judiciously deploy IRS in different IRS-aided radar sensing systems to achieve the optimal sensing performance is a crucial problem. Specifically, for IRS semi-passive and active sensing, IRS should be deployed at a location with a clear IRS-target LoS path, as the measurements of radar sensing associated with the LoS path are directly related to the target's location. This thus gives rise to a fundamental trade-off between increasing the IRS altitude to enhance the LoS probability and lowering the IRS altitude to reduce the path loss with ground target, as illustrated in Fig. \ref{deployment}(a).
On the other hand, for IRS passive sensing, the target's direction-of-arrival (DoA) cannot be resolved if there is only one single LoS path between the BS and IRS. This is because it requires at least two degrees-of-freedom to identify both the complex path gain and the angle parameter for target localization. Therefore, for IRS passive sensing, it is desirable to place the IRS at a location possessing a strong LoS path with the target, as well as an adequate number of NLoS paths with the BS.

Next, we consider the IRS deployment design for the general multi-IRS aided radar sensing system in a more complex radio propagation environment. In Fig. \ref{deployment}(b), we illustrate two typical IRS deployment strategies which are applicable to all the three IRS-aided radar sensing architectures: 1) \emph{centralized IRS deployment}, where all reflecting elements are aggregated to form a single large-size IRS, and 2) \emph{distributed IRS deployment}, where the reflecting elements are divided to form multiple small-size IRSs, each located near one region of interest. The centralized IRS deployment strategy is more advantageous for offering higher beamforming gain; while in contrast, the distributed strategy is more likely to establish LoS paths with the target in high-blockage environment. Hence, IRS placement and elements allocation should be jointly designed to balance the above trade-off for wireless sensing, which deserves future investigation.

Furthermore, in high-mobility scenarios (e.g., highway and railway), the channels between the static BS and in-vehicle users change rapidly due to the
environment's random scattering, which makes radar sensing more challenging to implement. In this case, IRS can be properly installed on the surface of high-speed vehicles or along the roadside to aid radar sensing, referred to as the \emph{vehicle-side} IRS and \emph{road-side} IRS, respectively, as illustrated in Fig. \ref{deployment}(c). Specifically, for vehicle-side IRS, as the users largely remain quasi-static with the IRS, the IRS can be exploited to track the relatively low-mobility in-vehicle users, while the BS can easily keep track of the large-size IRS even in high mobility. However, IRS needs to be separately deployed in each vehicle for enhancing its radar sensing performance, hence resulting in high hardware costs for vehicle manufacturing. Alternatively, deploying roadside IRSs for vehicle sensing can help exploit the potential multi-IRS cooperative localization gain for improving the radar sensing performance. The design and performance comparison of the above two IRS-aided sensing systems in high-mobility scenarios have not been studied yet, and deserve further investigation.

\subsection{IRS Channel Acquisition}
For IRS-aided radar sensing, channel acquisition refers to obtaining useful channel information for extracting target's physical parameters (e.g., DoA) embedded in the multi-reflection channel. Different from the conventional radar sensing system with only one unknown BS-target channel, the IRS-aided radar sensing system involves not only the unknown channel directly associated with the target (e.g., IRS elements$\rightarrow$target), but also other target-unrelated unknown channels (e.g., BS$\rightarrow$IRS elements channel in IRS passive/semi-passive sensing, and IRS controller$\rightarrow$IRS elements channel in IRS active sensing). As such, the static BS/IRS controller$\rightarrow$IRS elements channel can be estimated in an offline phase, followed by an online phase that utilizes such estimated channel information to extract the target's DoA embedded in the IRS elements$\rightarrow$target channel (more details will be given in Section II-E).

There are various efficient methods for estimating the BS/IRS controller-IRS elements channel for the three IRS-aided radar sensing architectures. For IRS semi-passive/active sensing, the IRS elements can be switched to the OFF mode, while the installed sensors are activated to receive the
signals from the BS/IRS controller for estimating their associated
channels to IRS elements. On the other hand, for IRS passive sensing, anchor nodes can be deployed near the IRS for facilitating its associated BS to acquire the BS-IRS channel.

\subsection{IRS Reflection Design}
Different from IRS-aided communication that usually aims to maximize the received SNR/signal-to-interference-plus-noise ratio (SINR) for rate maximization, IRS-aided radar sensing considers the performance metrics directly related to parameter estimation and/or target detection. Specifically, for parameter estimation,
Cram\'er-Rao lower bound (CRLB) or mean squared error (MSE) minimization is usually adopted.
Since CRLB is generally a complicated function of the received SNR \cite{liuan}, CRLB minimization may not be equivalently reduced to the more convenient but less accurate SNR maximization (albeit this may hold in some cases of radar sensing without IRS \cite{crb}).
To balance complexity and performance,
for IRS active sensing, the authors in \cite{shao_target} proposed to optimize IRS's reflective beamforming over time by maximizing the average received signal power at IRS sensors. Without any prior information, the resulting optimal IRS reflection design over time was shown to be an omnidirectional beampattern in the angular domain for scanning unknown targets in all possible directions. On the other hand, if given prior information on the region of target location, IRS reflection can be more delicately devised to form a flat and wide beampattern uniformly covering the specific region of interest, thus increasing the received SNR for target localization.

Besides the optimization-based methods, the codebook-based IRS reflection design can also be explored to reduce the design complexity. For DoA estimation, the number of scanning beams formed by IRS should be large enough to cover the sensing area of interest. However, given the sum-/average-power constraint, this method can only support a small transmit power per beam. Alternatively, the hierarchical codebook-based IRS reflection design can be applied to balance the above tradeoff. Specifically, IRS wide beams can be generated in the first phase to scan the entire area to determine the sector where the target locates; then, in the second phase, IRS can steer narrow beams within this sector to further resolve the fine-grained target DoA \cite{RSPjoint}.

Next, for IRS passive/semi-passive sensing, the IRS reflection, in general, needs to be jointly designed with the BS transmit beamforming to optimize the sensing performance, which renders the optimization problem more difficult to solve.
To address this issue, the alternating optimization technique can be employed to sub-optimally solve this problem by iteratively optimizing one of the BS transmit beamforming and IRS reflection with the other being fixed \cite{fangjun}. To further reduce the computational complexity, one can design the BS transmit beamformer so that it points towards the IRS controller, while IRS reflections over time can be dynamically tuned to find the target DoA with respect to the IRS controller. This design significantly differs from the conventional MIMO radar, where the BS directly sends orthogonal waveforms to scan the targeted area. In general, such a joint design of BS transmit beamforming and IRS reflection should be superior to the IRS reflection design alone in terms of DoA estimation accuracy, which requires further research, especially for the case without any prior information of the target location.

\subsection{Sensing Algorithm}
Given the IRS reflection or beam scanning design, the BS/IRS sensors receive a sequence of signals reflected by the target over time, which are then used for target detection (i.e., determining the presence or absence of a target or target parameter estimation). Specifically, for IRS active/semi-passive sensing, the echoes from different targets usually have different directionality at the IRS sensors (i.e., target$\rightarrow$IRS sensors angle). Thus, the detection/estimation techniques adopted in conventional MIMO radar are also applicable to IRS active/semi-passive sensing. For instance, the celebrated subspace-based methods, such as the multiple signal classification (MUSIC) algorithm and estimation of signal parameters via rotational invariance techniques (ESPRIT), can be employed to estimate the target DoA with super-resolution.

However, for IRS passive sensing, the echoes from different targets and then reflected by the IRS hit the BS antennas in the same direction (i.e., IRS$\rightarrow$BS angle), thus making the simultaneous detection/estimation of multiple targets more difficult.
In this case, the computationally efficient ESPRIT algorithm cannot be directly applied to DoA estimation, since the necessary condition (i.e., the array displacement invariance structure depends directly on the target location) for using it may not hold. Alternatively, the non-subspace-based methods, such as the maximum likelihood estimation (MLE), can be applied to estimate the target DoA in IRS passive sensing, which, however, incurs prohibitively high complexity. As such, how to design a low-complexity DoA estimation algorithm for IRS passive sensing is a new and challenging problem to solve in future work.
For multi-target detection, instead of distinguishing different targets based on their echo signals' directions, one can also detect the transmit signature sequences by establishing a one-to-one mapping between signature sequence and target direction to improve the detection performance \cite{qingmulti}.

Although deploying multiple IRSs can provide more opportunities to establish LoS paths with targets, collecting the global information at all IRSs for cooperative localization may incur unaffordable high overhead and complexity. One promising approach to tackle this issue is by developing efficient distributed sensing methods, where each IRS acts as a local DoA estimator, and distributed IRSs exchange  low-dimensional local information with
each other, thus avoiding the need of a global data-fusion processing.
However, this approach may incur notable performance loss due to incomplete data
collection. To further improve the performance and yet maintain low complexity, tensor-based algorithms can be applied to exploit the inherent Katri-Rao structure of multi-dimensional signals.

\section{Numerical Results}
In this section, we present numerical results to compare the sensing performance of different IRS-aided radar sensing architectures as well as sensing algorithms. Specifically, we assume that the IRS has $128$ reflecting elements and $8$ sensors, while the BS has $128$ transmit antennas and $8$ receive antennas. The BS-IRS distance is $100$ m, and the IRS controller-IRS elements distance is $0.5$ m. We assume that all channels are LoS. The discrete Fourier transform (DFT)-based IRS passive reflection design proposed in \cite{shao_target} is adopted for all IRS-aided sensing architectures.
\begin{figure}[t!]
\setlength{\abovecaptionskip}{-0.cm}
\setlength{\belowcaptionskip}{0.cm}
  \centering
\fbox{\includegraphics [width=0.43\textwidth] {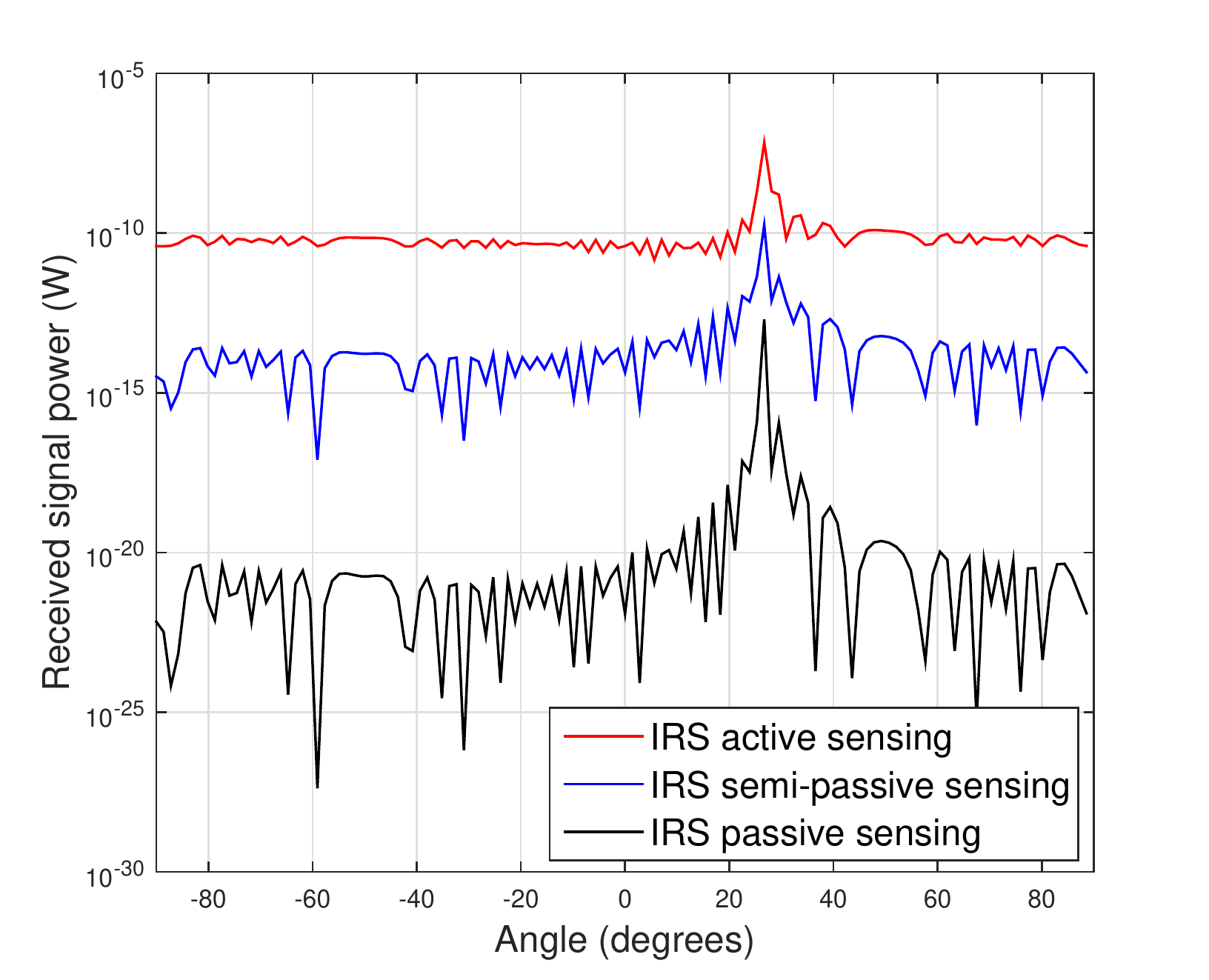}}
\caption{Beampattern comparison of three IRS-aided radar sensing architectures.}
\label{signalpower}
\end{figure}

In Fig. \ref{signalpower}, we compare the IRS beampatterns of different IRS-aided radar sensing architectures. We assume that the target is located at $30^{\circ}$ horizontally from the IRS. It is observed that the beampatterns of three architectures all correctly focus their mainlobes towards the target direction. In addition, IRS active sensing is observed to achieve a significantly larger received signal power than IRS semi-passive/passive sensing at the target angle. This is because IRS active sensing not only has the minimum path loss, but also exploits both the IRS passive beamforming gain and the direct echo link gain, as discussed in Section II-A. In contrast, IRS semi-passive sensing has a lower received signal power since the signal traveling distance from the transmitter to the IRS is much longer. Moreover, IRS passive sensing has the least received signal power due to the highest path loss.

\begin{figure}[t!]
\setlength{\abovecaptionskip}{-0.cm}
\setlength{\belowcaptionskip}{0.cm}
  \centering
\fbox{\includegraphics [width=0.43\textwidth] {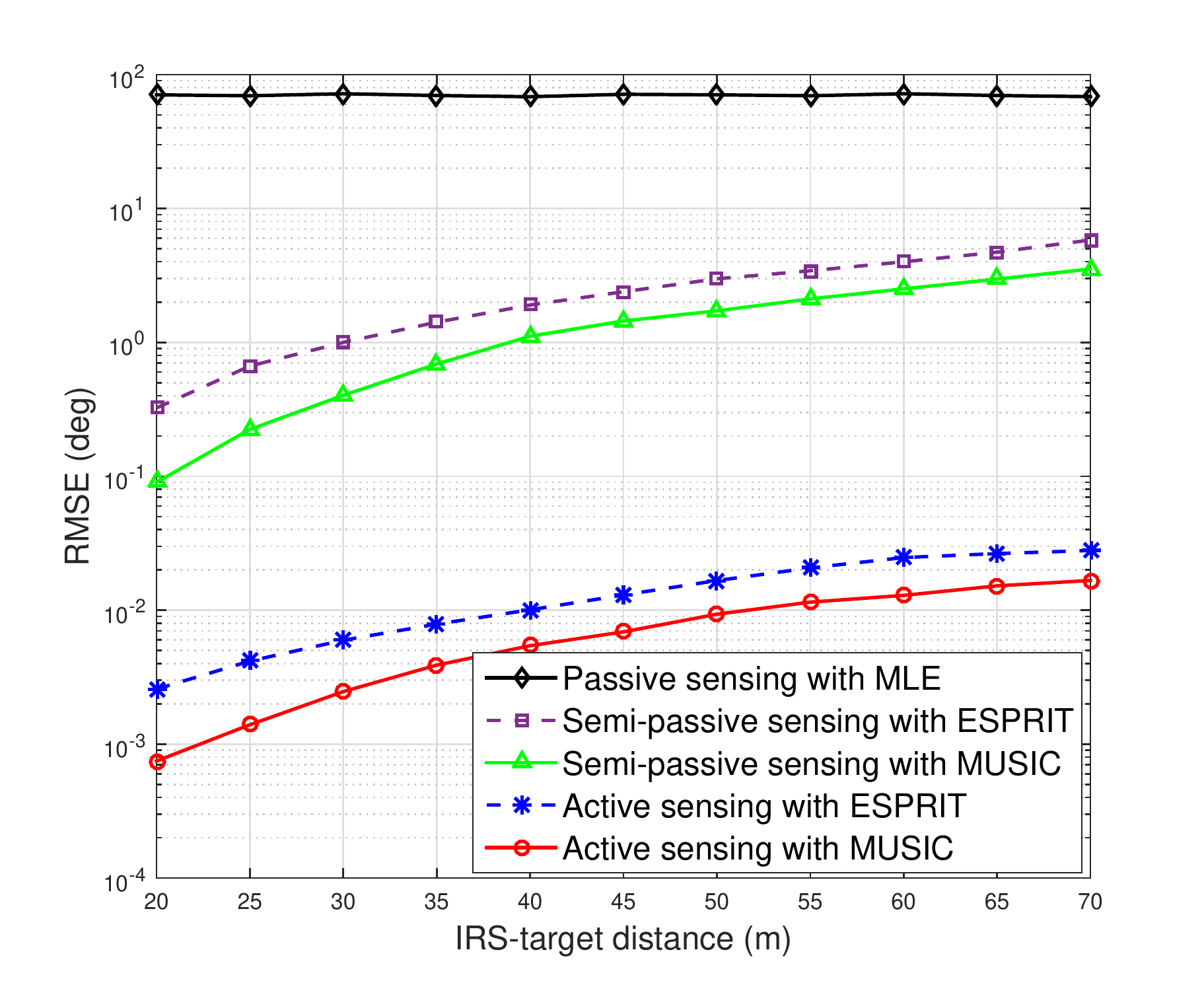}}
\caption{RMSE versus IRS-targets distance for different IRS-aided radar sensing architectures and sensing algorithms.}
\label{multiuser}
\end{figure}

Next, in Fig. \ref{multiuser}, we compare the root mean square error (RMSE) performance of different IRS-aided radar sensing architectures with different sensing algorithms, including MUSIC, ESPRIT, and MLE. Two targets are placed at the horizontal angles of $60^{\circ}$ and $65^{\circ}$, respectively. First, it is observed that for either the ESPRIT or MUSIC algorithm,
IRS active sensing consistently achieves a much smaller RMSE than IRS passive/semi-passive sensing under different IRS-targets distances. It is worth noting that the targets' DoAs cannot be resolved by IRS passive sensing, since this requires that there are at least two signal paths in the BS-IRS channel, which is not true in this case with LoS channel only. Second, one can observe that MUSIC achieves a smaller RMSE than ESPRIT, while the former generally requires higher computational complexity due to the need of exhaustive search over all possible directions.

\section{Other Extensions}
In addition to the above design issues for IRS-aided wireless sensing, there also
exist other related and important problems that
are open and worth investigating in future work. In the following, we outline such extendable directions to motivate future work.
\subsection{Non-terrestrial IRS Aided Sensing}
Besides terrestrial IRS, the non-terrestrial IRS can also be employed to enable high-accuracy sensing cost-effectively. For instance, a UAV-mounted IRS can be exploited to assist the BS to ``see"
the terrestrial targets from different angles by creating multiple LoS links to bypass terrestrial obstacles. However, Doppler shift due to UAV mobility may cause ambiguity in the velocity estimation. Therefore, the joint design of UAV position/trajectory, IRS reflection, and terrestrial BS beamforming/sensing is crucial to achieve optimal sensing performance. Besides, for low earth orbit (LEO) satellites enabled target tracking, their high velocity brings the requirement for steerable antennas. One effective approach is to deploy IRSs on the backside of solar panels of LEO satellite for realizing dynamic passive beamforming/beam scanning, thus its ground coverage area can be adaptively adjusted for enabling more efficient terrestrial communication and sensing.

\subsection{Active-IRS-Aided Sensing}
To compensate for the severe product-distance path loss, active IRS has been recently proposed, which is able to achieve simultaneous signal reflection and amplification at moderately higher hardware and energy cost than the conventional passive IRS \cite{changactive}.
Instead of using massive number of passive reflecting
elements, a smaller-size active IRS with much less number of active reflecting units can be deployed to effectively amplify the reflected signal, thus leading to enhanced signal power at the target for improving the sensing performance. However, active IRS also introduces non-negligible amplification noise, thus calling for efficient reflection designs to balance the tradeoff between the signal and noise amplification for wireless sensing.

\subsection{IRS for ISAC}
IRS for ISAC is a promising direction, since IRS is able to create virtual LoS links for improving both the communication and sensing performance. In particular, IRS reflection should be carefully designed to balance the communication and sensing performance. One simple approach is by using orthogonal time sharing, where IRS-aided communication and sensing are executed in orthogonal time with carefully designed time allocation. Besides, by properly deploying IRSs in the networks, the IRS reflection can be exploited to simultaneously improve the communication and sensing channel gains, if the two channels are similar or highly correlated. Moreover, how to design an efficient IRS-aided ISAC system without significantly changing the existing protocols in cellular/WiFi systems and achieve the optimum communication-and-sensing performance tradeoff are important directions for future research.

\section{Conclusions}
In this article, we propose IRS-aided wireless sensing to achieve cost-effective performance enhancement. Specifically, we discuss the new opportunities of applying  IRS to overcome the performance limitations of existing wireless sensing systems. Moreover, we present three practical IRS-aided radar sensing architectures and compare their performance and implementation cost/complexity, as well as the pertinent design issues and promising solution approaches. In particular, we show that IRS active sensing can achieve the best sensing performance compared to IRS passive/semi-passive sensing. Open problems and other extendable research directions for IRS-aided wireless sensing are also outlined to inspire future work.
This article is expected to provide a useful and effective guide for future research on IRS-aided wireless sensing.

{\small{\section*{Biographies}
\noindent {\bf Xiaodan Shao} [M'22] (shaoxiaodan@zju.edu.cn) received her Ph.D. degree in 2022 from Zhejiang University. She is with The Chinese University of Hong Kong, Shenzhen, China.
\newline

\noindent {\bf Changsheng You} [M'19] (youcs@sustech.edu.cn) received his Ph.D. degree in 2018 from The University of Hong Kong. He is currently an Assistant Professor with Southern University of Science and Technology, China.
\newline

\noindent {\bf Rui Zhang} [F'17] (elezhang@nus.edu.sg) received his Ph.D.
degree from Stanford University in 2007. He is now an X.Q. Deng Presidential Chair Professor with the School of Science and Engineering, The Chinese University of Hong Kong, Shenzhen, China. He is also with the ECE Department of National University of Singapore, Singapore.}}


\begin{thebibliography}{1}
\bibitem{6gisac}
C. De Lima \emph{et al}., ``Convergent communication, sensing and localization in 6G systems: An overview of technologies, opportunities and challenges," \emph{IEEE Access}, vol. 9, no. 1, pp. 26902-26925, Jan. 2021.

\bibitem{ISAC1}
D. K. Pin Tan \emph{et al}., ``Integrated sensing and communication in 6G: Motivations, use cases, requirements, challenges and future directions," in \emph{Proc. IEEE Int. Online Symp. Joint Commun. Sens.}, Mar. 2021, pp. 1-6.

\bibitem{liuan}
A. Liu, \emph{et al}., ``A survey on fundamental limits of integrated sensing and communication," \emph{IEEE Commun. Surv. Tut.}, vol. 24, no. 2, pp. 994-1034, Feb. 2022.

\bibitem{imag}
D. Huang, R. Nandakumar, and S. Gollakota, ``Feasibility and limits of Wi-Fi imaging", in \emph{Proc. ACM Conf. Emb. Net. Sensor Sys.}, Nov. 2014, pp. 266-279.

\bibitem{qingqing}
Q. Wu, S. Zhang, B. Zheng, C. You, and R. Zhang, ``Intelligent reflecting surface-aided wireless communications: A tutorial," \emph{IEEE Trans. Commun.}, vol. 69, no. 5, pp. 3313-3351, May 2021.

\bibitem{overview}
C. Pan, \emph{et al}., ``An overview of signal processing techniques for RIS/IRS-aided wireless systems," \emph{IEEE J. Sel. Topics Signal Process.}, vol. 16, no. 5, pp. 883-917, Aug. 2022.

\bibitem{RSPjoint}
R. S. Prasobh Sankar, B. Deepak, and S. P. Chepuri, ``Joint communication and radar sensing with reconfigurable intelligent surfaces," in \emph{Proc. IEEE Int. Workshop Signal Process. Adv. Wireless Commun. (SPAWC)}, Nov. 2021, pp. 471-475.

\bibitem{radarin}
Z. Jiang \emph{et al}., ``Intelligent reflecting surface aided dual-function radar and communication system," \emph{IEEE Syst. J.}, vol. 16, no. 1, pp. 475-486, Mar. 2022.

\bibitem{song}
Z. Yang \emph{et al}., ``MetaSLAM: Wireless simultaneous localization and mapping using reconfigurable intelligent surfaces," \emph{IEEE Trans. Wireless Commun.}, early access, Oct. 17, 2022.

\bibitem{specBei}
S. Lin, B. Zheng, F. Chen, and R. Zhang, ``Intelligent reflecting surface-aided spectrum sensing for cognitive radio," \emph{IEEE Wireless Commun. Lett.}, vol. 11, no. 5, pp. 928-932, May 2022.

\bibitem{shao_target}
X. Shao, C. You, W. Ma, X. Chen, and R. Zhang, ``Target sensing with
intelligent reflecting surface: Architecture and performance," \emph{IEEE J. Sel. Areas Commun.}, vol. 40, no. 7, pp. 2070-2084, Jul. 2022.

\bibitem{fangjun}
F. Wang, H. Li, and J. Fang, ``Joint active and passive beamforming for IRS-assisted radar," \emph{IEEE Signal Process. Lett.}, vol. 29, no.1, pp. 349-353, Dec. 2022.

\bibitem{crb}
F. Liu, Y. Liu, A. Li, C. Masouros, and Y. C. Eldar, ``Cram\'er-Rao bound optimization for joint radar-communication beamforming," \emph{IEEE Trans. Signal Process.}, vol. 70, no. 1, pp. 240-253, Dec. 2021.


\bibitem{qingmulti}
K. Meng, Q. Wu, R. Schober, and W. Chen, ``Intelligent reflecting surface enabled multi-target sensing," \emph{IEEE Trans. Commun.}, vol. 70, no. 12, pp. 8313-8330, Dec. 2022.

\bibitem{changactive}
C. You and R. Zhang, ``Wireless communication aided by intelligent reflecting surface: Active or passive?," \emph{IEEE Wireless Commun. Let.}, vol. 10, no. 12, pp. 2659-2663, Dec. 2021.
\end{thebibliography}
\end{document}